\documentclass[11pt,twoside]{article}
\usepackage{asp2010}

\resetcounters

\begin{document}

\title{Using Absorptive Linear Polarization Spectroscopy 
to Understand Imbedded Stars}
\author{J. R. Kuhn, B. Geiss, and D. M. Harrington
\affil{Institute for Astronomy, University of Hawaii, \\
34 Ohia Ku St, Pukalani, Maui, HI 96790, USA}
}

\begin{abstract}
Sensitive measurements of the linearly polarized spectra of stars can
be used to deduce geometric properties of their otherwise unresolved
circumstellar environments. This paper describes some of the evidence for optical pumping  and absorptive linear 
polarization and explores some
interesting applications of linear spectropolarimetry for obtaining
spatial information from imbedded stars.
\end{abstract}

\section{Introduction}
Linear spectropolarimetry of stars obtained 
with polarization sensitivity of
$0.1\%$ or better often shows linear polarization variability across absorption
lines \citep[cf.][]{vin02,vin05b,har07,har09a,har09b}. This effect is observed over a broad range of stellar conditions and in many
lines in a stellar spectrum. We've shown \citep{kuh07} that the anisotropy of the
stellar radiation field can lead to anistropy in the ground-state atomic
energy levels responsible for the observed stellar absorption lines. This
quantum mechanical effect is called optical pumping (OP) in the
laboratory \citep{h72}, and has been known for some time to yield observable linear polarization effects in the solar spectrum \citep{t97}. 
Using detailed radiative models of Hydrogen
absorption above the surface of a star we've shown \citep{kuh07} that H$_\alpha$ is often quite sensitive to optical pumping. Our earlier models didn't
include the effects of hyperfine coupling and considered only the low-lying atomic levels in Hydrogen. 

Interpreting H$_\alpha$ polarization is complicated by the degenerate
hyperfine transitions that are not resolved in typical stellar
spectropolarimetry measurements. In this paper we'll outline possibilities
for simplifying the interpretation of the observed polarization 
across stellar absorption lines. For example, from certain non-hydrogenic 
lines we can show that optical pumping, not scattering polarization, is
the mechanism  responsible for producing spectropolarimetric 
variability. 

In stellar systems where the lines in the Balmer or Paschen sequence originate
from the same circumstellar region, we find further support for the OP
model by showing that the linear polarization ratios for the strongest
hydrogen lines can be computed from simple models. These polarization
ratios may be used to constrain the local circumstellar
environment. We examine several stellar systems where optical pumping in
the Balmer lines appears to originate from the same region and we
speculate on how detailed comparison of optical pumping in different
lines can be developed as a useful tool.

In stellar systems with a rotating disk we illustrate how
the absorptive linear polarization due to optical pumping can be used to
determine some basic geometric properties. Thus,
the orientation of a disk in the sky and its tilt with respect to the plane of the sky can be inferred from the absorptive linear polarization
line profiles. Finally we show how linear spectropolarimetric data can be
inverted to obtain spatially resolved ``images'' of the intervening gas near the imbedded star.

\section{Absorptive Polarization}

\begin{figure}[h!]
\plotone{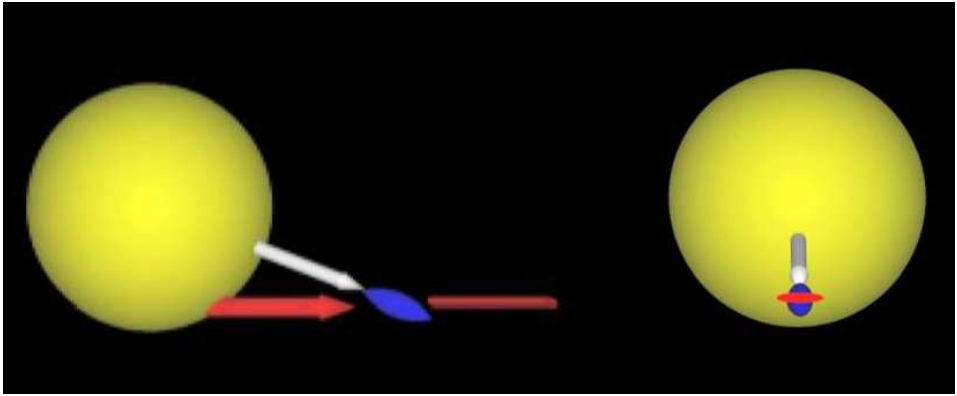}
\caption{Optical pumping geometry for a cloud (blue ellipse) observed along
a line-of-sight (red arrow) and radiatively excited from anisotropic stellar
radiation (white arrow)}
\label{opsgeom}
\end{figure}

Optical pumping originates from the lower level of atomic transition levels where the 
equilibrium density of the magnetic quantum number states depends
on the magnetic quantum number $m$. For
example, if a lower-level state with J=1 has a different electron density
for the $m=0$, compared to the $m= -1, +1$ levels then these atoms will
yield a line opacity which depends on the incident light direction and
linear polarization states. The population anisotropy is
determined by the angular anisotropy of the incident radiation.
The optical pumping (OP) mechanism geometry we describe can be simplified to a ``pumping" beam (or light cone) represented by the mean direction 
of the radiation which
induces anisotropy in the atomic sublevels, and a ``probe" beam which is 
the direction of the optical radiation that is selectively 
absorbed by the pumped atomic gas in the line-of-sight to the observer. 

Figure~\ref{opsgeom} shows schematically how OP works in the
circumstellar environment of a star. A classical analogy is indicated
in this figure. We consider an absorptive ``cloud'' between the observer
and a star indicated by the blue ellipse. The pumping radiation originates
along the direction of the white arrow and effectively polarizes the dipole
response of the atomic system to lie along this direction, indicated schematically
here by the elongated blue ellipse. Light which is absorbed by the cloud
originates from the photosphere of the star along the line-of-sight direction
to the observer (the red arrow). The absorbed radiation induces an electronic
dipole response in the cloud which favors a direction along the elliptical long axis. Thus, for initially unpolarized radiation it is the probe beam electric field direction most
closely aligned with the ellipse
which is selectively absorbed and scattered. The perpendicular
electric field orientation is transmitted with lower opacity. The red
ellipse projected in the side panel shows the enhanced polarization direction
of the transmitted light as seen from the observers perspective.
In general OP absorptive polarization yields
a transmitted beam polarized perpendicular to the plane containing the 
pumping and probe beam directions, although a negative polarization can
be induced in some atomic transitions as we show below.

\subsection{Database}

Our group has amassed a considerable database of precision linear
spectropolarimetry measurements from a broad range 
(about 100) of ``imbedded'' stars.
This includes HAeBe's, Post-AGB, RV-Tau's, Wolf-Rayet, and others \citep{har07,har08,har09a,har09b}. 
Part of this
database has been obtained with our own HiVIS spectrograph on the 
Haleakala AEOS 3.7m telescope but it also includes data obtained from
the ESPaDOnS instrument on the CFHT. The data described below was
obtained with the CFHT and comes primarily from archival data records.
These data were obtained with spectral resolution R$>$50000 and
with typical 2$\sigma$ linear spectropolarimetric accuracy of 0.1\% in the line
profiles.

\subsection{Non-scattering Lines}

Various mechanisms have been proposed to describe the linear spectropolarimetric
variability seen in stellar absorption lines \citep[cf.][]{vin05a,mcl79,oud99}. For example, scattering
polarization from free electrons or resonant line scattering 
is often invoked \citep{mcl79}. Unpolarized line scattering in combination with 
polarized Thomson scattering can also yield polarization variability across spectral lines \citep{mcl79}.
Scattering polarization requires an atomic upper level which allows
for a population anisotropy. This occurs if the net angular
momentum (including nuclear spin coupling) yields a total angular momentum
greater than 1/2. Calcium is relatively abundant with no net nuclear spin.
The CaII triplet transitions at 849.8, 854.2, and 866.2nm are an
excellent multiplet to test scattering as a polarization mechanism
since the lines have similar formation properties and the longest
wavelength term cannot be polarized by scattering. \citet{man03}
studied these transitions in the solar
spectrum in order to explain observed solar spectral polarization.

The 866.2nm transition upper level is a J=1/2, $^3P_{1/2}$
state, while the other transitions originate from the $^3P_{3/2}$ level. The
lower level in each of the triplets is a J=3/2 or 5/2 $^2D$ level. Thus if
line polarization is detected in the 866nm line it cannot be due to
scattering polarization. The angular momentum of the lower levels in all of
these transitions allows for optical pumping. Thus, by detecting linear polarization in the reddest of the CaII near-IR triplet lines, we have clear
evidence of OP absorptive polarization which cannot
be caused by resonant scattering effects.

\subsection{CaII NIR Triplet results}

\begin{figure}[h!]
\plotone{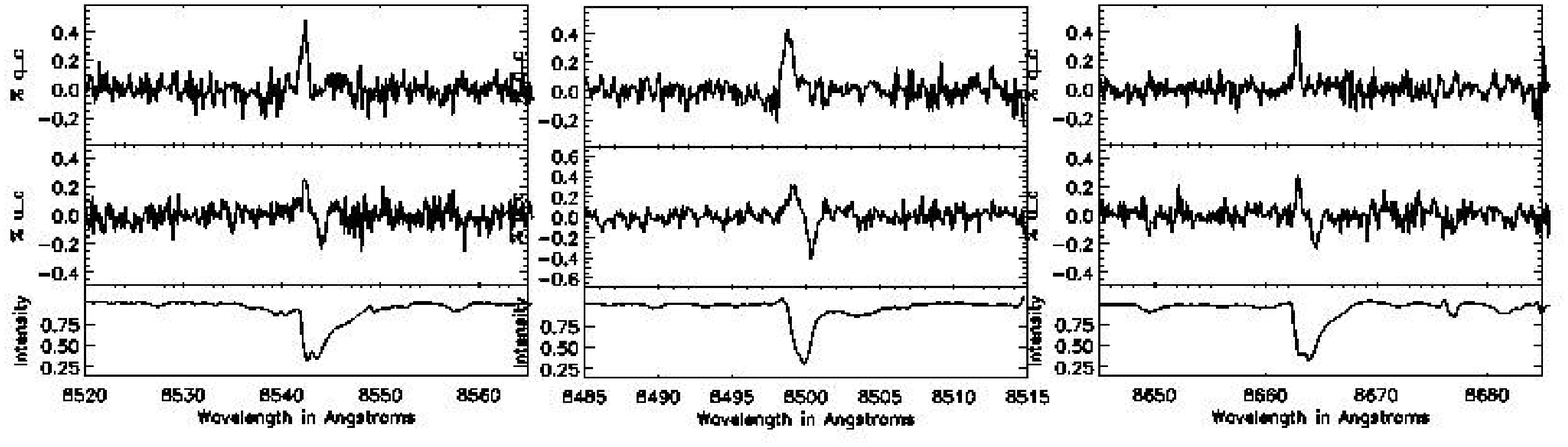}
\caption{Spectra and continuum normalized Stokes Q and U spectra
for U Mon in the CaII triplet lines}
\label{caumon}
\end{figure}

The CaII triplet lines in many of our stars are in emission, but of those
systems with these lines in absorption, about half show complex Stokes Q,U
polarization structure. Invariably the linear polarization morphology
across the red triplet component is similar to the two bluer triplet
components. An example is illustrated here. U Mon is an RVTau 91d variable
star. Fig. ~\ref{caumon} shows the intensity spectra and continuum normalized polarized spectra.

The Stokes Q and U polarized spectra show what we now know to be
typical spectral line polarization variation, localized in the absorptive
part of (in this case) the CaII triplet line cores. 
The linear polarization
exhibits a typical amplitude of a few 0.1's of percent. The correspondence
between the 866.2nm and the bluer triplet polarized spectra is clear evidence
that it is an optical pumping mechanism which is responsible for 
the observed absorptive polarization signal. 
 
\subsection{Optical Pumping and Hyperfine Levels in Hydrogen}

\begin{figure}[h!]
\includegraphics[scale=0.5,angle=90]{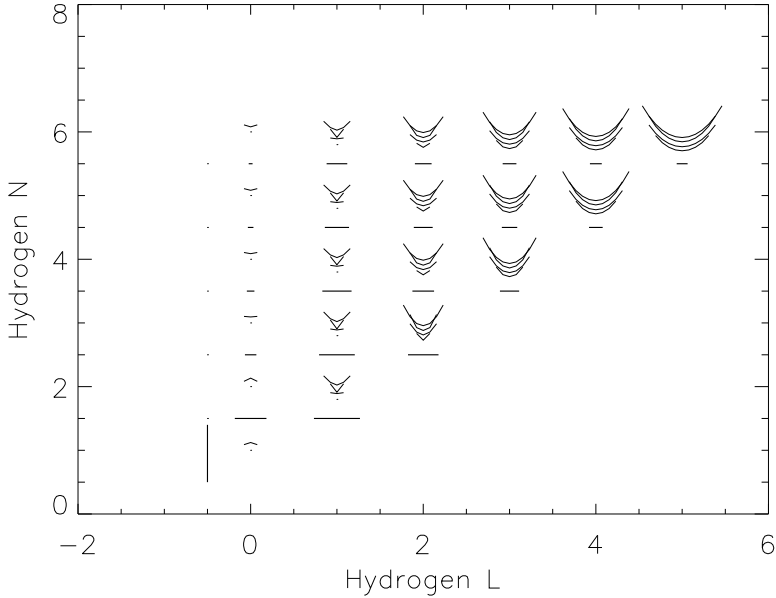}
\caption{The relative population of hydrogen levels for a 10,000K black-body
illuminating a collisionless gas with a 73deg cone angle. Hyperfine coupling
is included and the curved lines show the relative population from -m to +m
where m is the maximum magnetic total spin quantum number. The plot shows
this population versus principal quantum number n and orbital angular momentum
L. Thus for each n,L there are four levels with F=L+1, 2$\times$L, 
and L-1. The thicker
horizontal lines under each group of 4 total spin levels shows the relative
population versus L for given n. The thick vertical lines are proportional to
the relative population of all levels at given quantum number n. For example,
here a 10000K black-body leaves most of the electrons in n=1 states. Only
the L=0,1 states have significant population at larger n.}
\label{term10000_73}
\end{figure}

Hyperfine coupling of the atomic nuclear spin to the electronic levels implies
that even J=0 ground-state levels may be sensitive to anisotropy in the
radiation environment of a star. We've included this effect in hydrogen
and built a radiation coupled model that includes 364 levels and
about 5000 transitions up to $n=6$ in hydrogen. Following \citep{kuh07}
we've solved for the m-dependent equilibrium population in each of these
levels as a function of the background radiation field anisotropy (parametrized
by the radiation ``cone-angle'') and the radiation temperature. We assume
that atomic collisions and magnetic field quantum coherence
effects can be neglected. For the sufficiently intense radiation field and
low density stellar environments that we consider here these are generally good
assumptions. Thus Fig~\ref{term10000_73} shows that a 10,000K black-body
with a 73deg illumination angle produces higher populations in the largest
m-states at larger L. Note that the n=1, L=0 ground-state has 
opposite m-dependent
anisotropy, that is, the m=0 populations are greater than higher magnitude
m states. The relative population of higher n and L states is small
but for more intense radiation 
(higher temperature or larger illumination angle) the larger n and L states are
relatively more heavily populated. 
As the cone angle (isotropy) increases the m-dependent
population anisotropy decreases.

A strong m-dependent density dependence in radiation coupled levels in an
atom can yield {\it opposite} population anisotropy in the upper and lower
transition states. This is clear in Fig.~\ref{term10000_73} where the low
lying L=0 states have higher m=0 populations, while the L=1 states have
{\it lower} m=0 populations relative to higher magnitude magnetic quantum
numbers. In this
figure a positive or negative anisotropy is indicated by states with
line curvature up versus down. A
consequence of this is that as the radiative coupling changes because of
a stronger radiation field (pumping region is closer to the star), or
hotter radiation, the OP polarization changes rapidly. Thus,
changing anisotropy (as radiation illumination cone
angle changes) affects a competition between the opposite anisotropy composite
states to produce significant net OP line polarization changes.

\begin{figure}[h!]
\includegraphics[scale=.4,angle=90]{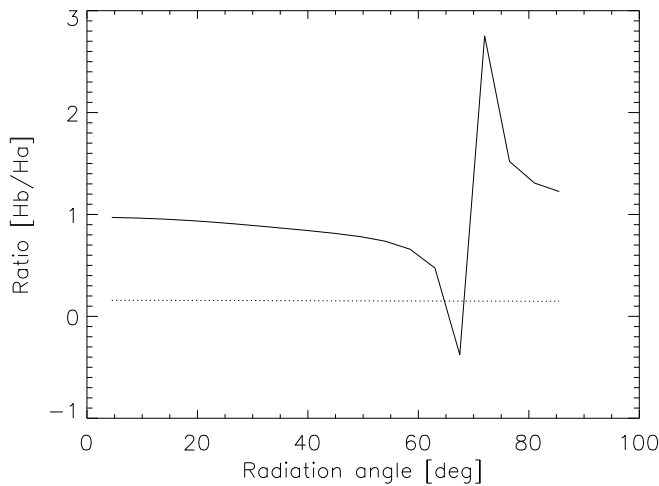}
\caption{The solid line shows the ratio of H$_\beta$/H$_\alpha$ polarization
and the dotted line shows the corresponding line intensity ratio versus
radiation illumination cone angle. Competition between opposite density
anisotropy low-lying hydrogenic levels causes a complicated polarization
ratio as radiation coupling varies with cone angle.}
\label{fig:rad10000}
\end{figure}

Polarization {\it ratios} of lines from the same term can be a very sensitive
probe of the radiation strength and isotropy due to the opposite density
anistropy that is possible in low-lying atomic states. For example, we have
computed the ratio of the theoretical Balmer line polarization in H$_\beta$/H$_\alpha$. Figure~\ref{fig:rad10000} plots the ratio of the
n=4 to n=2 and the n=3 to n=2 transition polarizations versus the
radiation illumination angle in degrees for a 10000K blackbody. The dotted
curve shows that while the line intensity ratio is nearly constant as the
radiation isotropy and strength vary, the polarization ratio even changes
sign in a complex way. Line polarization ratios are a sensitive probe of
the near stellar environment and contain radiation information distinct
from the intensity line ratios.

\begin{figure}[h!]
\plotone{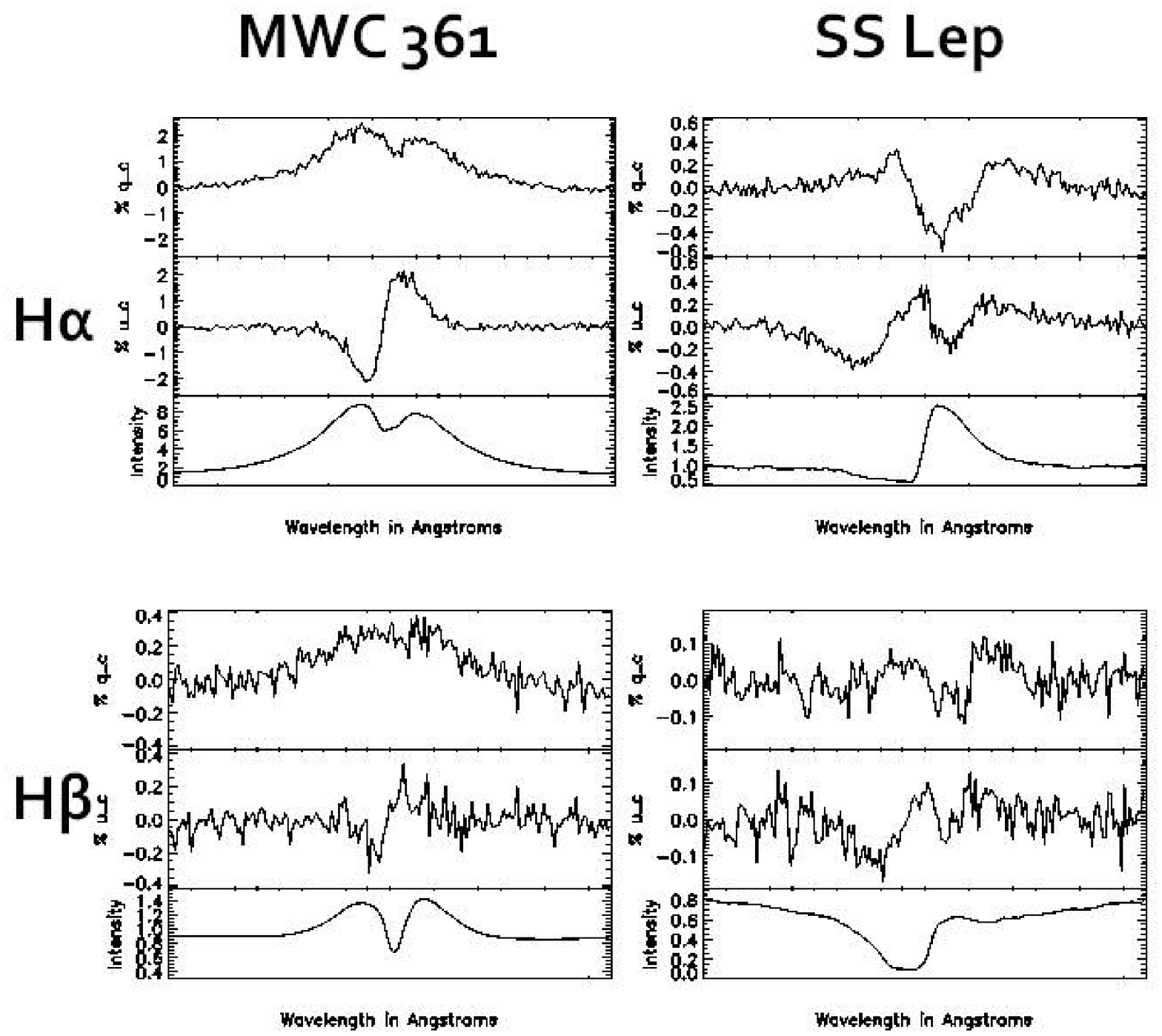}
\caption{Each panel shows the Q, U, and I spectra near H$_\alpha$
and H$_\beta$ forMWC 361 and SS Lep.}
\label{fig:umonmwcpol}
\end{figure}

Two examples illustrating a range of polarization and line ratios are SS Lep and
MWC 361. SS Lep is a symbiotic binary star and MWC361 is a Herbig Be star with a directly detected disk \citep{oka09}.
Figure~\ref{fig:umonmwcpol} shows H$_\alpha$ and H$_\beta$ for
these two stars. The intensity ratio of the absorptive component of SS Lep
and MWC 361 are about 0.3 and 0.2, while the corresponding polarization
ratios are 0.3 and 0.1. Our work to uniquely interpret such line and polarization
ratios is only beginning.

\subsection{Geometric Properties from Absorptive Linear Spectropolarimetry}

The mean radiation pumping direction at any point in the circumstellar cloud
defines a direction projected onto the plane of the sky that, in the simplest
approximation (see Fig.~\ref{opsgeom}), 
is perpendicular to the mean electric field 
polarization of the transmitted light through the cloud at that location. If
each location in the cloud encodes a definite doppler shift velocity, 
for example if
the cloud is actually a rotating disk projected against the stellar disk,
then the shape of the polarized absorptive
line spectrum with wavelength gives information on the position angle of the
disk in the sky and its inclination angle toward the observer.

\begin{figure}[h!]
\plotone{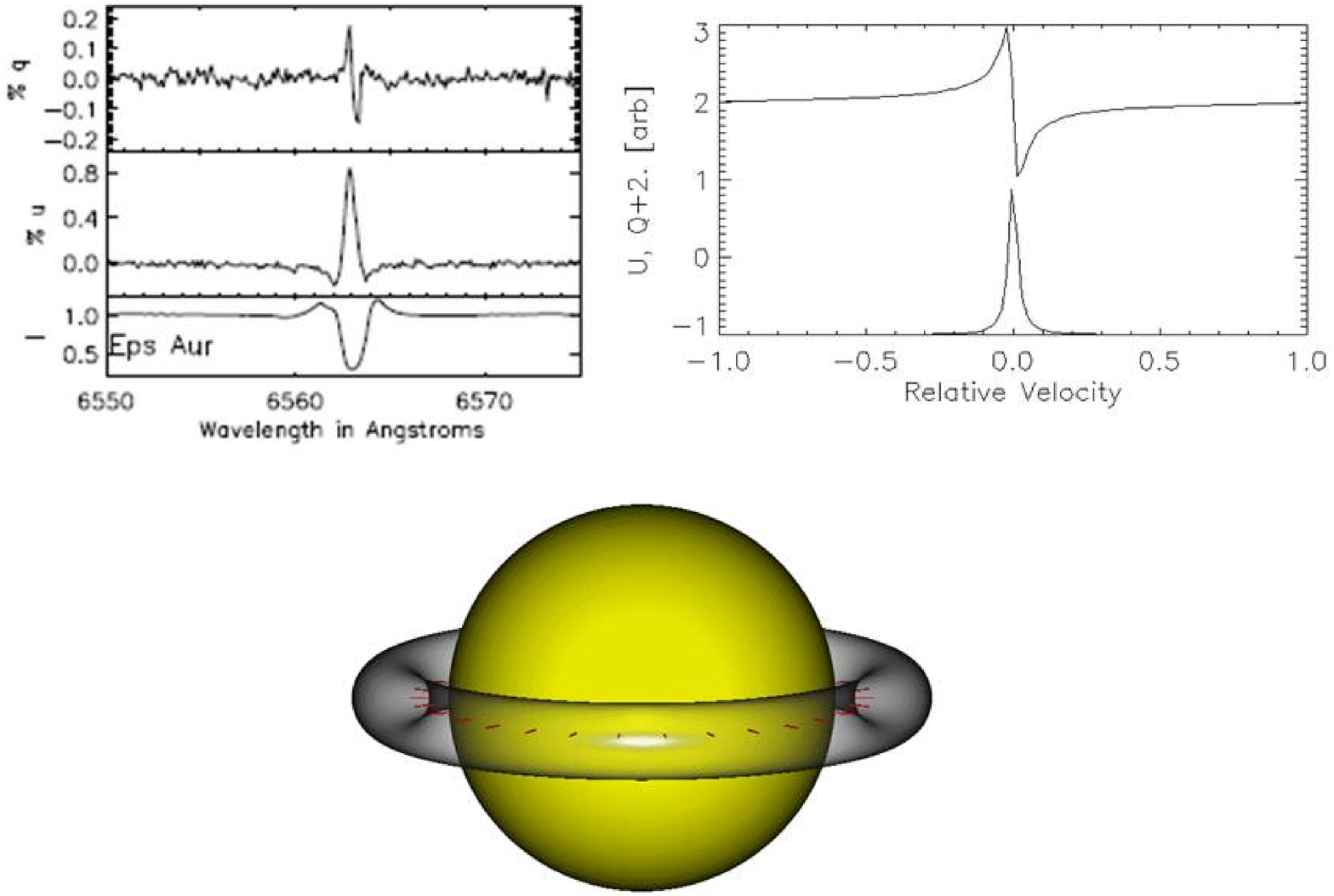}
\caption{The observed Q and U spectra for $\epsilon$ Aur and a
modeled linear polarization line profile for the indicated disk
geometry. The small red lines in the disk are perpendicular to the 
transmitted light linear polarization direction at that projected 
disk location.}
\label{smallangletilt}
\end{figure}

\begin{figure}[h!]
\plotone{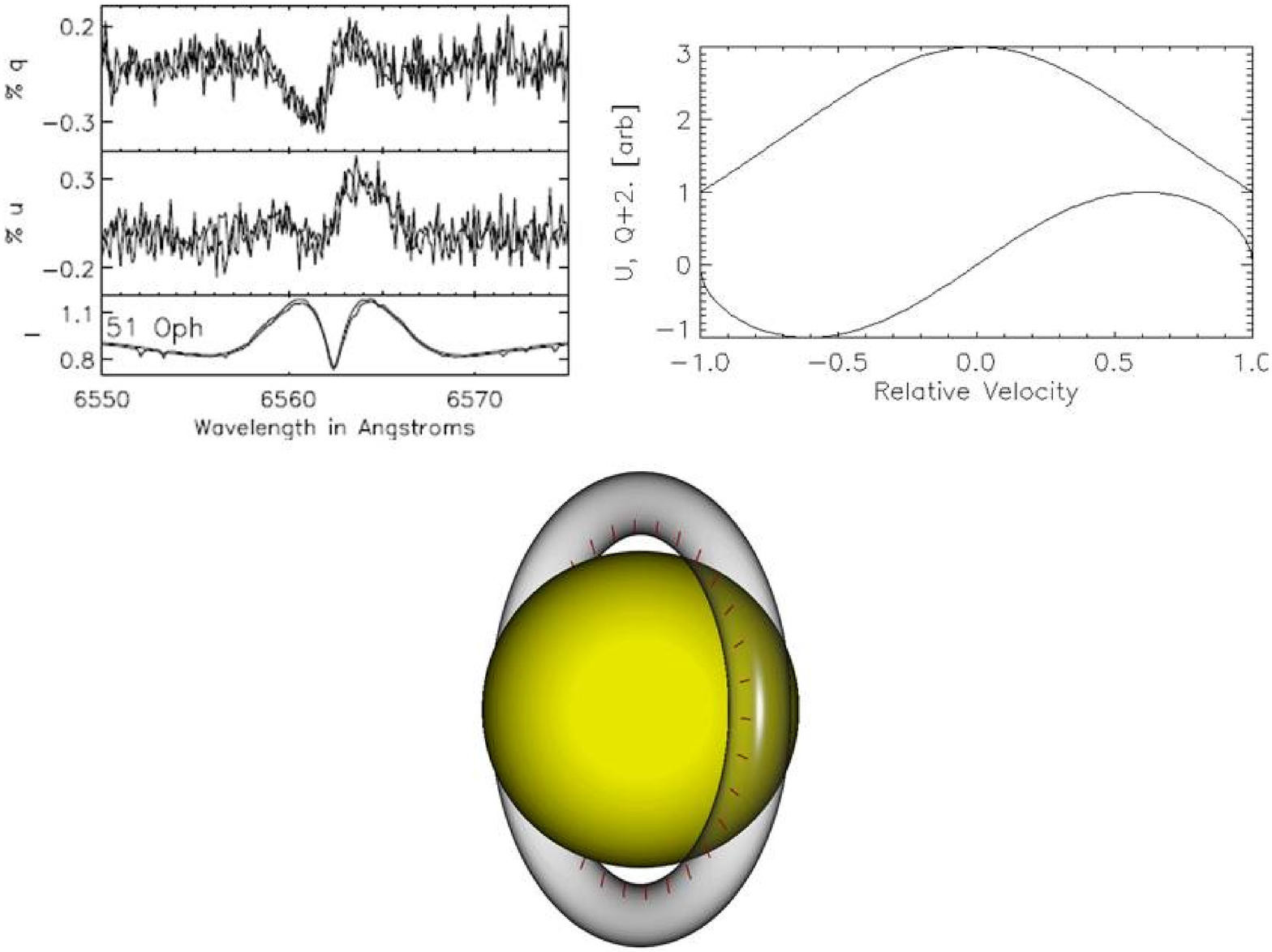}
\caption{The observed Q and U spectra for the star 51 Oph and a correspondingly
modeled linear polarization line profile for the indicated disk
geometry. The small red lines in the disk/ring figure panel are 
perpendicular to the 
transmitted light linear polarization direction at that location.}
\label{largeangletilt}
\end{figure}

Figure~\ref{smallangletilt} shows an example of the expected Q, U polarization
for a disk oriented as in the figure and observations from $\epsilon$ Aur that
are consistent with this orientation. Similarly Figure~\ref{largeangletilt} show
the likely orientation of 51 Oph derived from H$_\alpha$ linear absorptive
spectropolarimetry. 

\begin{figure}[h!]
\includegraphics[scale=.3]{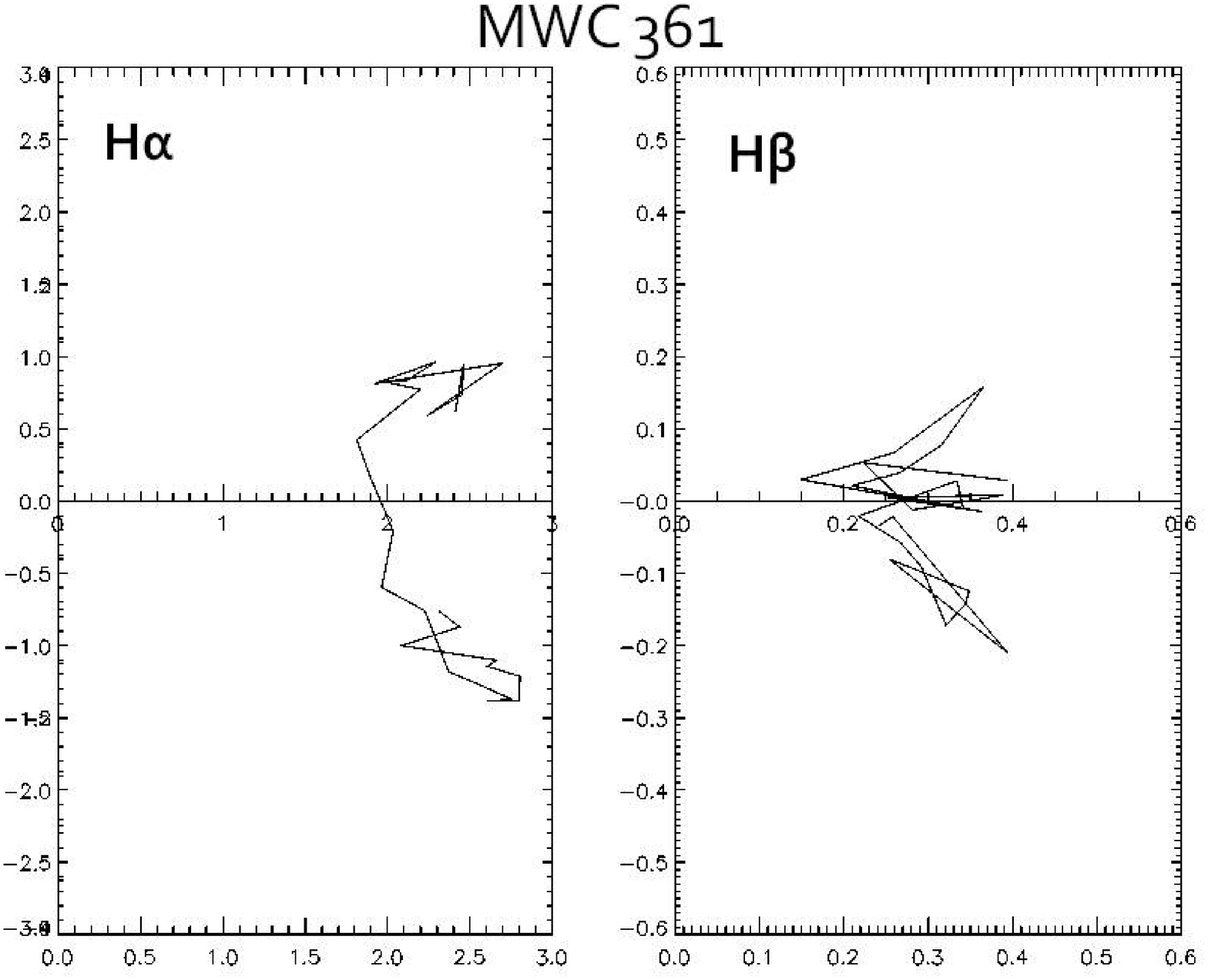}
\caption{The H$_\alpha$ and H$_\beta$ Q, U amplitude and direction
are plotted as a collection of points across the absorptive line profile.
Rotating these points to a symmetry line around the x-axis shows that the
disk position angle is about 5 degree in the sky.}
\label{MWCorient}
\end{figure}

For a star with sufficient symmetry in its disk a straightforward procedure to
derive the disk orientation is to plot the polarization angle at each wavelength
across the absorptive line profile. Thus, in the case of MWC 361, if we
derive an angle from $\theta = 1/2 arctan(Q/U)$ and a polarization amplitude
from $p=\sqrt{Q^2 + U^2}$ then a plot of $p(\theta )$ looks like Figure~\ref{MWCorient}. The orientation of the disk is found by rotating the 
plot until symmetry about the x-axis is achieved.  The data in Fig.~\ref{MWCorient} have been rotated about 5 degrees to yield symmetry
around the x-axis suggesting that the position angle of the MWC 361 disk
is 5 deg. This angle is consistent with the extended IR disk seen
by \citet{oka09} but is at odds with the modeled rotation axis obtained
from Zeeman Doppler Imaging \citep[cf.][]{oka09}.

\subsection{Absorptive Linear Polarization Imaging}

\begin{figure}[h!]
\includegraphics[scale=0.4]{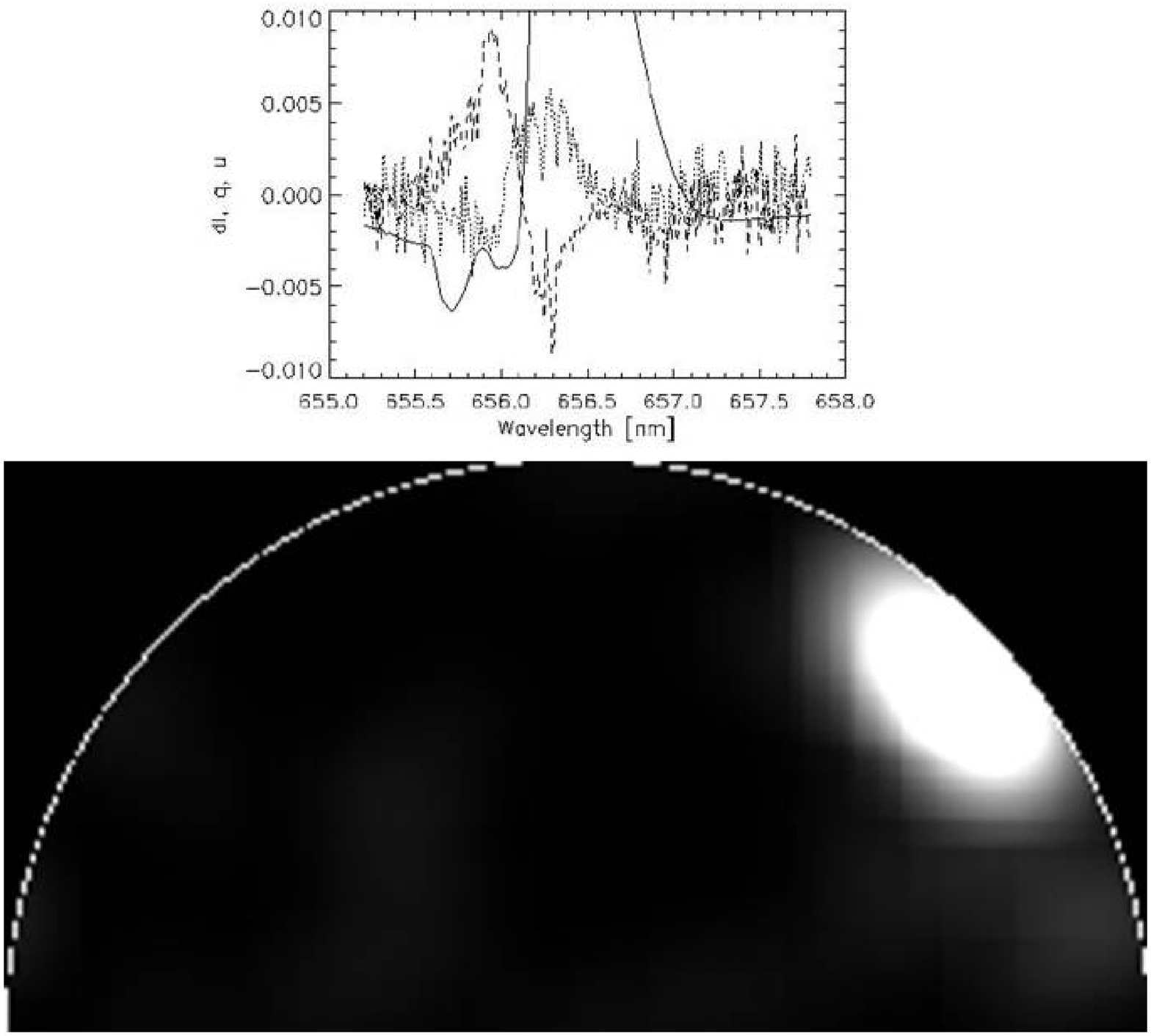}
\caption{From the high resolution I, Q, and U spectra of H$_\alpha$
we derive the projected optical depth of intervening hydrogen gas in
the side panel of this figure. Lighter shades indicate higher depth.}
\label{abaurinversion}
\end{figure}

With sufficient velocity resolution across the absorptive 
line profile the intensity and
polarization amplitude and Q, U spectra may be inverted to yield the 
column density projected onto the star. Because the linear polarization is
localized to the line-of-sight to the star, and analogous to the case of
Doppler Zeeman Imaging for a stellar magnetic field, there is spatial
information about the structure of the obscuring gas in front of the
star in the Q, U spectra. In general the OP signal
originates close to the pumping source -- Kuhn et al. (2007) argued this
is likely to be within 1 stellar radius of the photosphere. Either from
the polarization amplitude or the absorbed intensity at each wavelength we
derive a local cloud ``density'' or optical depth. At each wavelength the
Q, U polarization ratio defines an angle on the disk and the doppler
velocity defines a surface (based on an assumed 3-space rotation profile or
a wind velocity) the obscuring material must lie along. 
Alternatively the local polarization amplitude yields
the separation between the local cloud position and the geometric
center of the illuminating photosphere. We can demonstrate this using 
the H$_\alpha$ spectropolarimetry for AB Aur in order to 
obtain an ``image'' of the intervening cloud as shown
in Figure~\ref{abaurinversion}. Here a radial outflow 
wind velocity was assumed and the
inversion has a 180 degree ambiguity because of the
linear polarization 2-fold rotational symmetry. 
The light shades in this figure indicate where the
largest absorption against the stellar disk occurs as obtained from a single
epoch inversion using high resolution I, Q, and U H$_\alpha$ spectra of
AB Auriga.

\section{Conclusions}

Absorptive linear polarization is a nearly ubiquitous phenomenon in
the spectra of imbedded stars. It is often visible in polarized Stokes Q and U
spectra when a sensitivity better than 1\% is achieved. We've shown
that  it is not a scattering phenomenon and is well described by optical
pumping due to ground-state
anisotropy from the  anisotropic stellar radiation environment. 
We've shown here that there is new information on the
stellar environment contained in linear spectropolarimetry that is
distinct from, for example, line intensity ratios. 
With simple assumptions, the geometry
of stellar disks can be inferred. With additional assumptions
we've demonstrated how
high resolution spectropolarimetric data can be inverted to
yield ``images'' of the intervening gas from polarized 
absorptive line profiles.

\acknowledgements This research has been supported by the Institute for
Astronomy on Maui. We're grateful to Svetlana Berdyugina who contributed
to the hydrogenic hyperfine calculations.

\end{document}